\documentclass[aps,prb,twocolumn,10pt,longbibliography,superscriptaddress,floatfix]{revtex4-2}
\pdfoutput=1
\usepackage{amsmath}
\usepackage{amsfonts}
\usepackage{amssymb}
\usepackage{mathtools}
\usepackage[colorlinks=true,citecolor=blue,linkcolor=blue,urlcolor=blue]{hyperref}
\usepackage{bm}
\usepackage{times}
\usepackage{cases}
\usepackage{wasysym}
\usepackage[version=4]{mhchem}
\usepackage{graphicx}
\usepackage{subfigure}
\usepackage{booktabs}  
\usepackage{multirow}
\usepackage{tikz}
\usepackage[normalem]{ulem}
\usepackage{mhchem}
\usepackage{physics}
\usepackage{comment}
 
 \usepackage{physics}
 
\begin{document}
\title{Flavor Nernst effects in quantum paramagnets}

\author{Bowen Lu}
\thanks{These authors contributed equally to this work.}
\affiliation{State Key Laboratory of Surface Physics and Department of Physics, Fudan University, Shanghai 200433, China}
\affiliation{International Center for Quantum Materials, School of Physics, Peking University, Beijing 100871, China}

\author{Bowen Ma}
\thanks{These authors contributed equally to this work.}
\email{bowenphy@hku.hk}
\affiliation{Department of Physics and HK Institute of Quantum Science \& Technology, The University of Hong Kong, Pokfulam Road, Hong Kong, China}

\author{Yue Yu}
\affiliation{State Key Laboratory of Surface Physics and Department of Physics, Fudan University, Shanghai 200433, China}

\author{Gang Chen}
\affiliation{International Center for Quantum Materials, School of Physics, Peking University, Beijing 100871, China}

\date{\today}
    
\begin{abstract}
Recent advances in spin transport research have highlighted the potential of quantum paramagnets as platforms for exploring novel phenomena and developing next-generation technologies. In this paper, we investigate the flavor Nernst effect (FNE) in quantum paramagnets, focusing on the Hall-type thermal spin transport of crystal electric field (CEF) excitations with spin-orbit couplings. As a proof of principle, we investigate the quantum paramagnetic ground state in an effective spin-1 Hamiltonian with Dzyaloshinskii-Moriya interactions and a large hard-axis anisotropy. We employ linear flavor-wave theory to analyze the low-energy excitations, and obtain the flavor Nernst coefficients from the linear response theory. We demonstrate the FNE in a 2D pyrochlore thin film with an all-in-all-out Ising axis configuration, and investigate their dependence on temperature, anisotropy, DM interaction, and external fields. Our results reveal the connection between the FNE and the Berry curvature of the CEF excitations, suggesting potential applications in manipulating thermal spin currents and exploring topological spin transport phenomena in quantum paramagnets.
\end{abstract}

\maketitle
\section{Introduction}
Quantum paramagnets, characterized by the absence of long-range magnetic order down to zero temperature, often arise in materials where strong quantum fluctuations suppress conventional magnetic ordering~\cite{knolle2019Field}. In many such systems, the interplay between spin-orbit coupling (SOC) and crystal electric field (CEF) effects leads to a complex local Hilbert space with multiple energy levels~\cite{rau2016SpinOrbit}. These CEF excitations can be viewed as generalized triplets, analogous to the triplon excitations in dimerized magnets~\cite{akbari2023Topological,mcclarty2017Topological,romhanyi2015Hall}.

As these relevant excitations in quantum paramagnets are charge neutral, the traditional electrical approach usually fails here. In contrast, the thermally driven transport has emerged as a powerful probe for investigating the properties of elementary excitations in correlated quantum materials~\cite{zhang2024Thermal,onose2010Observation,katsura2010theory}. For example, the linear residual term of ultra-low temperature longitudinal thermal conductivity attributed to mobile fermionic excitations has been regarded as evidence of gapless quantum spin liquids~\cite{ni2019Absence,bourgeois-hope2019Thermal,zhu2023Fluctuating}. On the other hand, the transverse thermal Hall effect provides valuable information about the Berry curvature of the excitations~\cite{zhang2024Thermal,ma2024upper,boulanger2020Thermal}. Recent experiments have revealed significant thermal Hall signals in various quantum magnets, including \ce{Tb2Ti2O7}~\cite{li2013Phononglasslike}, \ce{Yb2Ti2O7}~\cite{tokiwa2016Possible}, and Cd-kapellasite \ce{CdCu3(OH)6(NO3)2 * H2O}~\cite{akazawa2020Thermal}, where CEF excitations play a crucial role. Besides carrying energy, these excitations can also transport spin information by thermally driven forces. Recently, the interplay between spin transport and heat currents has garnered significant attention in the field of condensed matter physics, and play a crucial role in the development of spintronics~\cite{zutic2004Spintronics,hoffmann2015Opportunities,barker2021Review,elahi2022Review,nakayama2021Aboveroomtemperature,back2019Special,uchida2021Spintronica}. Many thermal spintronic phenomena, such as spin Seebeck effects and spin-dependent Peltier effects, have been experimentally observed and theoretically explained~\cite{flipse2012Direct,bakker2010Interplay,maekawa2017spin,bhardwaj2018Spin,adachi2013Theory,uchida2016Thermoelectric,ma2020longitudinal}.  Among those studies, the topological transport of spin has emerged as a particularly intriguing area, offering insights into fundamental quantum phenomena and potential applications in next-generation devices~\cite{armitage2018Weyl,tokura2019Magnetic}. 
Similar to the thermal Hall effects, when the spin excitations acquire non-trivial band topology, a transverse spin current can be generated in response to a longitudinal temperature gradient and a perpendicular magnetic field. This is the so-called spin Nernst effect, and it has been realized by topological magnons in spin-orbit coupled magnets~\cite{cheng2016spin,zyuzin2016Magnon,meyer2017Observation,shiomi2017Experimental,sheng2017Spin,ma2021Intrinsic,zhang2022Perspective}. Obviously, this magnonic spin Nernst effect requires the ground state to be magnetically ordered, and thus the system temperature is restricted by the phase transition temperature $T_c$. However, as it is the thermally activated excitation that is carrying non-zero spin, the ground state in principle does not need to be in an ordered phase. In addition, $T_c$ is usually smaller than 80 K in many candidate topological magnets (e.g. two candidates with existing experiments for topological magnons, \ce{CrI3} and \ce{Lu2V2O7} has $T_c\approx 45$ K and $70$ K respectively~\cite{huang2017Layerdependent,onose2010Observation}), while the thermal Hall signal has been observed in quantum paramagnet \ce{Tb2Ti2O7} up to 142 K~\cite{hirschberger2015Thermal}. We are then motivated to extend the ideas of spin Nernst effect to these quantum paramagnets, where the CEF excitations have non-zero angular momenta~\cite{babkevich2015Neutrona,thalmeier2024Induced}. In comparison with the linear spin-wave theory in magnetic systems~\cite{kittel2018introduction}, the linear flavor-wave theory~\cite{joshi1999elementary,li19984} is applied to obtain the CEF excitations from a paramagnetic ground state, and thus we dub the spin Nernst effects in quantum paramagnets as flavor Nernst effects.

In the remainder of this paper, we provide details of the flavor Nernst effects in quantum paramagnets. In Sec.~\ref{sec:Quantum Paramagnets}, we formulate an effective spin-1 Hamiltonian to realize the quantum paramagnetic ground state and introduce the linear flavor-wave representation for the CEF excitations. In Sec.~\ref{sec:Linear response theory}, within the linear response theory, we show that the thermal response coefficients of flavor Nernst effects can be expressed by generalized spin Berry curvature, which is strongly connected to the band topology of the CEF excitations. In Sec.~\ref{sec:2D pyrochlore thin film}, we evaluate the flavor Nernst coefficients in a 2D pyrochlore thin film as an example, and study the effects of the system temperature, the coupling strength, and the external magnetic field on the flavor Nernst spin transport. And finally, we conclude the paper with some discussions and suggest possible directions for future work in Sec.~\ref{sec:Discussions}.

\section{Quantum Paramagnets}
\label{sec:Quantum Paramagnets}
\subsection{Model Hamiltonian}
To illustrate the idea of the quantum paramagnetic ground state and excited CEF states in a simple way, we consider a spin-1 system with the local moments from transition metal ions. For example, a 3d$^8$ electron configuration of Ni$^{2+}$ with octahedral crystal field, so that the lower $t_{2g}$ orbitals are fully occupied and the upper $e_g$ orbitals are half-filled, leading to an effective spin $S=1$~\cite{li2018competing}. For this case, there is no orbital degeneracy, and a pure spin Hamiltonian is adequate for describing the system, while in more general cases, orbital degrees of freedom may also need to be considered~\cite{kugel1982jahn,chen2010exotic}.

Due to the local ligand environment and the ligand-mediated hopping~\cite{pesin2010mott}, in addition to the usual Heisenberg exchange couplings $J$, we generically include the contribution from spin-orbit couplings as the antisymmetric Dyzaloshinskii-Moriya (DM) interaction~\cite{moriya1960anisotropic,dzyaloshinsky1958thermodynamic} $\bm{D}_{ij}$ between sites $ij$ and on-site single-ion anisotropy $\eta$. The spin Hamiltonian is then written as
\begin{align}
    H&=\sum_{\langle ij\rangle}\left(J \bm{S}_i\cdot\bm{S}_j+\bm{D}_{ij}\cdot\bm{S}_i\times\bm{S}_j\right)\nonumber\\
    &+\sum_i\left[\eta(\hat{\bm{z}}_i\cdot\bm{S}_i)^2-B(\hat{\bm{z}}\cdot \hat{\bm{z}_i})(\hat{\bm{z}}_i\cdot\bm{S}_i)\right],\label{H}
\end{align}
where $B$ is the Zeeman splitting from a magnetic field along the global $z$-direction $\hat{\bm{z}}$ and $\hat{\bm{z}}_i$ is the local Ising axis determined by the local ligand environment at site $i$~\cite{dun2016magnetic,dun2020quantum}.

Though a full magnetic phase diagram of the quantum ground state with varying interactions is difficult to obtain, for the purpose of this work, we focus on the large anisotropy limit with $\eta>0$, where the ground state should be a product of the singlet state $\ket{\hat{\bm{z}}_i\cdot\bm{S}_i=0}$ at each site $i$, i.e.,
\begin{align}
    |\text{quantum paramagnetic phase}\rangle\equiv\prod_{i}|\hat{\bm{z}}_i\cdot\bm{S}_i=0\rangle,
\end{align}
and the lowest several excited states are from $\hat{\bm{z}}_i\cdot\bm{S}_i=\pm 1$ with energy $\sim\eta\pm B$. Other exchange interactions, such as $J,\bm{D}_{ij}\ll\eta$, lead to the dispersive band structure of the excitations. Therefore, when the temperature $k_BT\gtrsim \eta\pm B$, the excited CEF states are thermally activated and the thermal transport of these excitations is then possible.

\subsection{Linear flavor-wave theory}

To obtain the dispersion of the low-energy excitations, we invoke the so-called linear flavor-wave theory~\cite{joshi1999elementary,li19984}.  Since the ground state is paramagnetic without ordering, the linear spin-wave theory with the Holstein-Primakoff expansion~\cite{holstein1940field} around the spin order cannot be applied here. Instead, we regard the states with $\hat{\bm{z}}_i\cdot\bm{S}_i=0,\pm 1$ (we denote $-1\equiv \bar{1}$ in the remaining) as three different flavors, and restrict ourselves to the Hilbert space spanned by these states $|f\rangle_i\equiv|\hat{\bm{z}}_i\cdot\bm{S}_i=f\rangle$ for each site $i$. These basis states form a representation of SU(3) with a set of generators $G_f^{f'}(i)=|f\rangle_i\langle f'|_i$. One can easily note that $G_f^{f'}(i)=\left[G_f'^{f}(i)\right]^\dagger$, and we further require a normalization condition that $1=G_0^{0}(i)+G_1^{1}(i)+G_{\bar{1}}^{\bar{1}}(i)$. 

The spin ladder operators can then be written as~\cite{ma2024upper}
\begin{align}
    S_i^+&\equiv(\hat{\bm{x}}_i+i\hat{\bm{y}}_i)\cdot\bm{S}_i\nonumber\\
    &=\sum_{ff'}\langle f|S_i^+|f' \rangle|f \rangle\langle f'|=\sqrt{2}\left[G_1^0(i)+G_0^{\bar{1}}(i)\right],\\
    S_i^-&\equiv(\hat{\bm{x}}_i-i\hat{\bm{y}}_i)\cdot\bm{S}_i\nonumber\\
    &=\sum_{ff'}\langle f|S_i^-|f' \rangle|f \rangle\langle f'|=\sqrt{2}\left[G_{\bar{1}}^0(i)+G_0^1(i)\right].
\end{align}
Similarly,
\begin{align}
     S_i^z&\equiv\hat{\bm{z}}_i\cdot\bm{S}_i=\sum_{ff'}\langle f|S_i^z|f' \rangle|f \rangle\langle f'|=G_1^1(i)-G_{\bar{1}}^{\bar{1}}(i),\\
    (S_i^z)^2&=\sum_{ff'}\langle f|(S_i^z)^2|f' \rangle|f \rangle\langle f'|=G_1^1(i)+G_{\bar{1}}^{\bar{1}}(i).
\end{align}
Here, $(\hat{\bm{x}}_i,\hat{\bm{y}}_i,\hat{\bm{z}}_i)$ spans a local coordinate for site $i$ 
as the flavors are defined by the local Ising axis $\hat{\bm{z}}_i$.

From the Young tableaux~\cite{kim2017linear}, the SU(3) algebra can be reproduced by two bosons $b$ and $\bar{b}$ as
\begin{align}
G_1^1(i)&=b^\dagger_ib_i,\\
G_{\bar{1}}^{\bar{1}}(i)&=\bar{b}^\dagger_i\bar{b}_i,\\
G_0^0(i)&=1-b^\dagger_ib_i-\bar{b}^\dagger_i\bar{b}_i,\\
G_{\bar{1}}^{1}(i)&=\bar{b}^\dagger_ib_i,\\
G_1^0(i)&=b^\dagger_i\sqrt{1-b^\dagger_ib_i-\bar{b}^\dagger_i\bar{b}_i}\approx b^\dagger_i,\\
G_{\bar{1}}^0(i)&=\bar{b}^\dagger_i\sqrt{1-b^\dagger_ib_i-\bar{b}^\dagger_i\bar{b}_i}\approx \bar{b}^\dagger_i,
\end{align}
and then we immediately obtain the linear flavor-wave representation of $\bm{S}_i$ as,
\begin{align}
    \left\{\begin{array}{lll}
     S_i^z&=b^\dagger_i b_i-\bar{b}^\dagger_i \bar{b}_i,\\
     S_i^-&=\sqrt{2}(\bar{b}^\dagger_i+b_i),\\
     S_i^+&=\sqrt{2}(\bar{b}_i+b^\dagger_i),
    \end{array}
    \right.
\end{align}
Physically, the bosonic operators $b^\dagger_i$ and $\bar{b}^\dagger_i$ ($b_i$ and $\bar{b}_i$) create (annihilate) an excitation with a ``magnetic flavor'' $\hat{\bm{z}}_i\cdot\bm{S}_i=+1$ and $\hat{\bm{z}}_i\cdot\bm{S}_i=-1$ from the quantum paramagnetic ground state respectively.

Generally, $\hat{\bm{z}}_i$ can orient differently for each site $i$, and thus the global spin U(1) symmetry is broken. This is similar to the non-collinear antiferromagnetic ordered phase, where the pairing of the bosonic operators does not vanish and a particle-hole doubling is necessary to give rise to a Bogoliubov–de Gennes (BdG) Hamiltonian~\cite{del2004quantum}. Therefore, in the momentum space, Hamiltonian Eq.~\eqref{H} with $m$-sublattices can be written in a particle-hole symmetric form as $H=\frac{1}{2}\sum_{\bm{k}}\bm{\Psi}_{\bm{k}}^\dagger H_{\bm{k}}\bm{\Psi}_{\bm{k}}$, where
\begin{gather}
    H_{\bm{k}}=
    \begin{pmatrix}
        A_{\bm{k}} & B_{\bm{k}}\\
        B^*_{-\bm{k}} & A^*_{-\bm{k}}
    \end{pmatrix},\label{eq:Ham}\\
    \bm{\Psi}_{\bm{k}}\!\!=\!\!\left(b_{1{\bm{k}}},\!\bar{b}_{1{\bm{k}}},...,b_{m{\bm{k}}},\!\bar{b}_{m{\bm{k}}},\!b^\dagger_{1,-{\bm{k}}},\!\bar{b}^\dagger_{1,-{\bm{k}}},...,b^\dagger_{m,-{\bm{k}}},\!\bar{b}^\dagger_{m,-{\bm{k}}}\right)^{\mathrm{T}}\!.
\end{gather}
The band dispersion of flavor-wave excitations can be determined as the positive eigenvalues of $\Sigma_z H_{\bm{k}}$, where $\Sigma_z=\begin{pmatrix}
    1 & 0\\
    0 & -1
\end{pmatrix}\otimes I_{2m}$ and $I_{2m}$ is the $2m\times 2m$ identity matrix. The details on the diagonalization can be found in the appendix~\ref{appendix:Bogoliubov–de Gennes Hamiltonian}.

\section{Linear response theory}
\label{sec:Linear response theory}

It has been shown that, with the DM interactions and/or the non-collinear Ising axes, the CEF excitations can be topologically non-trivial by breaking the time-reversal symmetry~\cite{ma2024upper}. With finite Berry curvature  $\bm{\Omega}_{n{\bm{k}}}$ from the $n$-th topological band, the wave-packet of the excitations will experience a transverse anomalous velocity as~\cite{xiao2010berry,cheng2016spin},
\begin{align}
    \dot{\bm{r}}_n=\frac{1}{\hbar}\frac{\partial E_{n{\bm{k}}}}{\partial {\bm{k}}}-\dot{{\bm{k}}}\times\bm{\Omega}_{n{\bm{k}}},
\end{align}
where $\bm{r}_n$ and $E_{n{\bm{k}}}$ is the packet center and the band dispersion of the $n$-th wavefunction. Since the CEF excitations additionally carry ``flavors'', similar to the quantum spin Hall effects of electrons~\cite{kane2005quantum} and spin Nernst effects of topological magnons~\cite{cheng2016spin,zyuzin2016Magnon}, if $\bm{\Omega}_{n{\bm{k}}}\neq 0$, we also expect a flavor Nernst effect (FNE), where a transverse spin current with polarization $s$ flowing along $\alpha$-direction can be induced by the longitudinal temperature gradient along $\beta$-direction as
\begin{align}
    j^s_\alpha=\nu^s_{\alpha\beta}\nabla_\beta T.
\end{align}
Here $\nu^s_{\alpha\beta}$ is the Nernst coefficient. 

By invoking the Heisenberg equation of motion~\cite{matsumoto2011theoretical,matsumoto2011rotational,li2020intrinsic}, the time evolution of the angular momentum operator $\tilde{\bm{S}}$ can be written as
\begin{align}
    \frac{\partial}{\partial t}\tilde{\bm{S}}=i[H,\tilde{\bm{S}}]=-\nabla\cdot\tilde{\bm{j}}+\tilde{\bm{T}},
    \label{eq:Heisenberg equation}
\end{align}
where $\tilde{\bm{S}}=\sum_{\bm{k}}\bm{\Psi}^\dagger_{\bm{k}}(S^x,S^y,S^z)\bm{\Psi}_{\bm{k}}$ with $S^{s}=\text{Diag}(\hat{\bm{s}}\cdot\hat{\bm{z}}_1,\dots,\hat{\bm{s}}\cdot\hat{\bm{z}}_m,\hat{\bm{s}}\cdot\hat{\bm{z}}_1,\dots,\hat{\bm{s}}\cdot\hat{\bm{z}}_m)\otimes
\begin{pmatrix}
    1 & 0\\
    0 & -1
\end{pmatrix}$, spin current operator $\tilde{\bm{j}}^s=\frac{1}{4}\sum_{\bm{k}}\bm{\Psi}_{\bm{k}}^\dagger\left[\bm{v}_{{\bm{k}}}\Sigma_z S^s+S^s\Sigma_z\bm{v}_{\bm{k}}\right]\bm{\Psi}_{\bm{k}}$ with $\bm{v}_{{\bm{k}}}=\nabla_{\bm{k}} H_{\bm{k}}$, and $\tilde{T}^s=-\frac{i}{2}\sum_{\bm{k}}\bm{\Psi}_{\bm{k}}^\dagger\left[S^s\Sigma_z H_{\bm{k}}-H_{\bm{k}}\Sigma_z S^s\right]\bm{\Psi}_{\bm{k}}$ acts as a spin torque operator. A spin current with polarization $s$, i.e. $\tilde{\bm{j}}^s$, is well-defined only if the spin-torque-like term $\tilde{T}^s=0$, or equivalently $[\Sigma_z\bm{S}^s, H]=0$. As we mentioned before, this conservation may not be satisfied with the general orientation of $\hat{\bm{z}}_i$'s, and the definition of spin current then seems to be problematic. However, it has been studied by linear response theory that the thermal response of the spin current can still be defined in an inversion symmetric system, where the contribution from $\tilde{\bm{T}}$ totally vanishes owing to the symmetry cancellation~\cite{li2020intrinsic}. The corresponding Nernst coefficient $\nu^s_{\alpha\beta}$ is expressed as

\begin{align}
    \nu^s_{\alpha\beta}=\frac{2k_B}{V}\sum_{n=1}^{2m}\sum_{{\bm{k}}}\tilde{\Omega}^s_{\alpha\beta,n{\bm{k}}}c_1[g(E_{n{\bm{k}}})],
    \label{eq:nu}
\end{align}
where $V$ is the size of the system, $c_1(x)=(1+x)\ln(1+x)-x\ln x$, $g(x)=(e^{x/k_B T}-1)^{-1}$ is the Bose-Einstein distribution, and $\tilde{\Omega}^s_{\alpha\beta,n{\bm{k}}}$ is a generalized spin Berry curvature~\cite{li2020intrinsic,ma2021Intrinsic} defined as
\begin{equation}
    \tilde{\Omega}^s_{\alpha\beta,n{\bm{k}}}=\sum_{n'\neq n}(\Sigma_z)_{nn}\frac{2\text{Im}[(j^s_{\alpha{\bm{k}}})_{nn'}(\Sigma_z)_{n'n'}(v_{\beta{\bm{k}}})_{n'n}]}{\left[(\Sigma_z)_{nn}E_{n{\bm{k}}}-(\Sigma_z)_{n'n'}E_{n'{\bm{k}}}\right]^2},
    \label{GsBC}
\end{equation}
where $(\dots)_{nn'}$ stands for $\langle\psi_{n{\bm{k}}}|(\dots)|\psi_{n'{\bm{k}}}\rangle$ with $|\psi_{n{\bm{k}}}\rangle$ the eigen-wavefunction for the $n$-th band such that $H_{\bm{k}}|\psi_{n{\bm{k}}}\rangle=E_{n{\bm{k}}}|\psi_{n{\bm{k}}}\rangle$, and Im[ ] extract the imaginary . In the collinear Ising axes case where $\hat{\bm{z}}_i=\hat{\bm{z}}$, we show in the appendix~\ref{appendix:Spin Berry curvature} that the spin Berry curvature reduces to the normal Berry curvature and Eq.~\eqref{eq:nu} recovers the form of spin Nernst effect studied in collinear antiferromagnets~\cite{cheng2016spin,zyuzin2016Magnon}.

\begin{figure}
    \centering
    \includegraphics[width=0.48\textwidth]{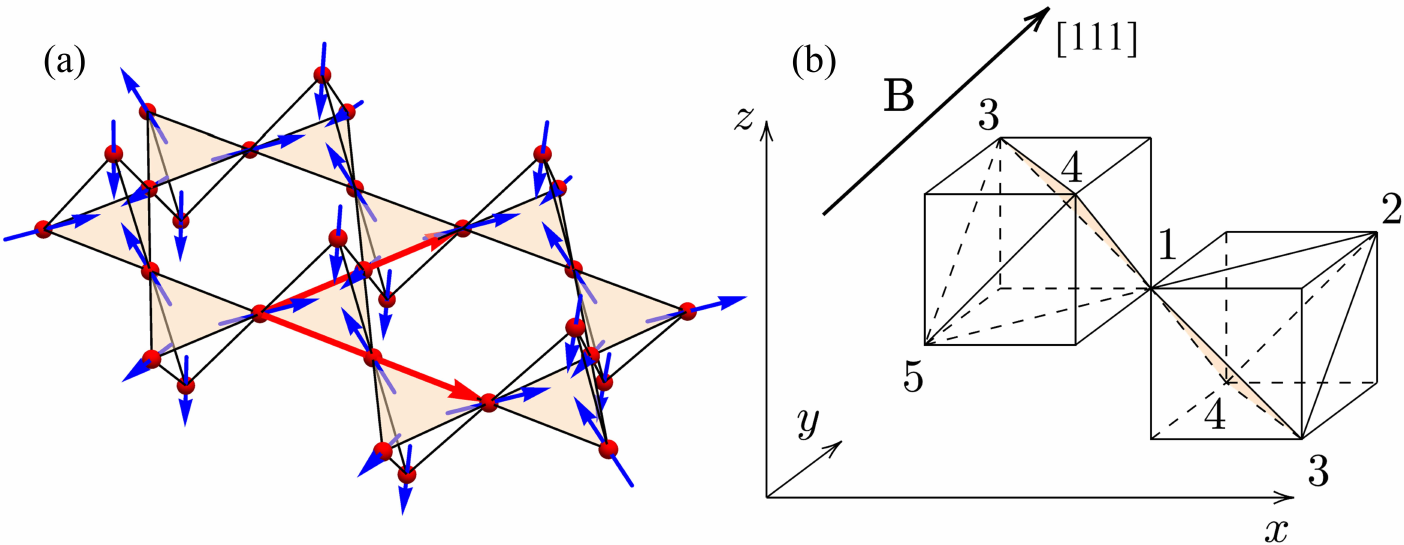}
    \caption{(a) The thin film pyrochlore lattice structure with AIAO Ising axes. Blue arrows show the local Ising axes on every site. The red arrows are the basis vectors $\bm{a}_1=a(1,0)$ and $\bm{a}_2=a(1/2,\sqrt{3}/2)$. (b) Label of every single site. The magnetic field $\bm{B}$ is along $[111]$ direction.}
    \label{fig:pyrochlore_lattice}
\end{figure}

\begin{table}
\caption{The local coordinate systems for the five sublattices.}
\begin{ruledtabular}
\begin{tabular}{cccccc}
$\mu$ & 1 & 2 & 3 & 4 & 5\\
$\hat{z}_\mu$ &$\frac{1}{\sqrt{3}}[11\bar{1}]$  &$\frac{1}{\sqrt{3}}[\bar{1}\bar{1}\bar{1}]$  &$\frac{1}{\sqrt{3}}[\bar{1}11]$  &$\frac{1}{\sqrt{3}}[1\bar{1}1]$  &$\frac{1}{\sqrt{3}}[\bar{1}\bar{1}\bar{1}]$\\
\end{tabular}
\end{ruledtabular}
\label{tab: local coordinate system}
\end{table}

\section{2D pyrochlore thin film}
\label{sec:2D pyrochlore thin film}
To study the flavor Nernst effect, we apply Eq.~\eqref{eq:nu} to a specific lattice model. One of the possible platforms to realize the quantum paramagnetic phase is the geometrically frustrated pyrochlore antiferromagnets, such as \ce{Gd2Pt2O7}~\cite{welch2022MagneticStructure}, \ce{NaCdCo2F7}~\cite{kancko2023StructuralSpinglass} and \ce{LiGaCr4O8}~\cite{he2021NeutronScattering}. For simplicity, instead of a 3D bulk system, we consider a 2D pyrochlore thin film grown along [111] direction as shown in Fig.~\ref{fig:pyrochlore_lattice}. It may be possible to realize this quasi-2D pyrochlore structure in real materials~\cite{liu2024chiralspinliquidlikestatepyrochlore}. One promising avenue is the fabrication of sandwiched heterostructures, where ultrathin layers of pyrochlore materials are grown between insulating spacer layers. Recent advances in molecular beam epitaxy and pulsed laser deposition techniques have enabled the synthesis of high-quality pyrochlore thin films with thicknesses down to a few unit cells~\cite{fujita2016AllinalloutMagnetic}. Researchers also developed a new situ film growth method "repeated rapid high-temperature synthesis epitaxy (RRHSE)"~\cite{kim2019InoperandoSpectroscopic,kim2020StrainEngineeringMagnetic} to improve the relaxed films problem.

\begin{figure}
\flushleft
    \includegraphics[width=0.46\textwidth]{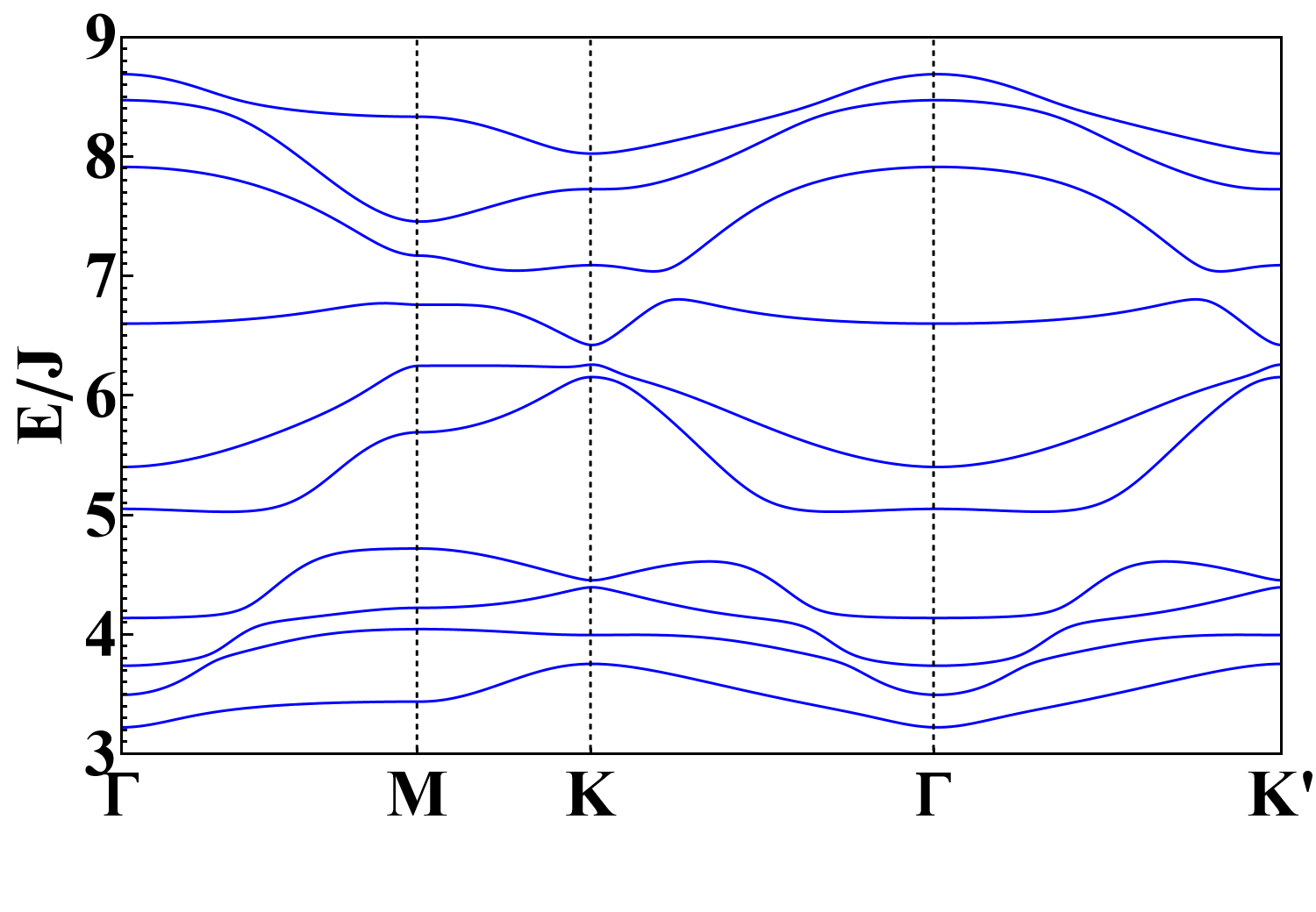}
    \caption{The band dispersion of the CEF excitations in the thin film pyrochlore from linear flavor-wave theory. Here, we set $\eta/J = 6$, $D/J=0.06$  and $B/J=0.6$.}
    \label{fig:dis_pyro}
\end{figure}

Along the [111] axis, the pyrochlore thin film consists of 5 sublattices forming alternating triangular and kagome layers~\cite{hu2012topological,laurell2017topological}. The Ising axes are in a so-called all-in-all-out (AIAO) configuration as the bulk system~\cite{li2018competing}, where $\hat{\bm{z}}_i$ is along the local [111] axis for each sublattice which is shown in Table~\ref{tab: local coordinate system}. The inversional symmetry is preserved so that the thermal spin current response Eq.~\eqref{eq:nu} is well-defined, while the bond inversional symmetry is broken, allowing non-zero DM interactions. The direction of the DM vectors can be determined by Moriya's rule~\cite{moriya1960anisotropic}. For the 12 bond in Fig.~\ref{fig:pyrochlore_lattice}, we have
\begin{align}
    \bm{D}_{12}=\frac{D}{\sqrt{2}}(-1,1,0)
\end{align}
and $\bm{D}_{ij}$ on other bands can be obtained from the lattice symmetry.

In the local coordinate basis, the spin Hamiltonian Eq.~\eqref{H} can be rewritten in a spin-ice-like expression~\cite{ross2011quantum} as
\begin{align}
    H=&\sum_{<ij>} [J_{zz} S^z_i S^z_j + J_{\pm}(S^+_i S^-_j+\text{H.c.})+ J_{\pm\pm}(\gamma_{ij}S^+_i S^+_j\nonumber \\
    &+ \gamma^*_{ij}S^-_i S^-_j)+ J_{z\pm}(\xi_{ij} S^z_iS^+_j + \xi_{ij}S^+_iS^z_j+\text{H.c.})]\nonumber\\
    &+ \sum_i (\eta(S^z_i)^2 - B S^z_i),
    \label{eq:Full_Ham}
\end{align}
where
\begin{align}
    J_{zz}&=\frac{1}{3}(2\sqrt{2}D-J),\quad J_\pm=-\frac{1}{6}(\sqrt{2}D+J),\nonumber\\
    J_{\pm\pm}&=-\frac{1}{3}(\frac{D}{\sqrt{2}}-J),\quad J_{z\pm}=\frac{1}{6}(D+2\sqrt{2}J),
\end{align}
and $\gamma_{ij}=-\xi^*_{ij}$ are bond-dependent phase variables. Taking $\gamma_{ij}$ as the $ij$-element of a matrix, one can express these variables in a $5\times 5$ matrix form as
\begin{equation}
    \gamma=
    \begin{pmatrix} 
        0               &1                   &e^{i2\pi/3}         &e^{-i2\pi/3}     &1\\
        1               &0                   &e^{-i2\pi/3}        &e^{i2\pi/3}      &0\\
        e^{i2\pi/3}     &e^{-i2\pi/3}        &0                   &1                &e^{-i2\pi/3}\\
        e^{-i2\pi/3}    &e^{i2\pi/3}         &1                   &0                &e^{i2\pi/3}\\
        1               &0                   &e^{-i2\pi/3}        &e^{i2\pi/3}      &0
    \end{pmatrix}.                        
\end{equation}

\begin{figure}[b]
    \flushright
    \includegraphics[width=0.48\textwidth]{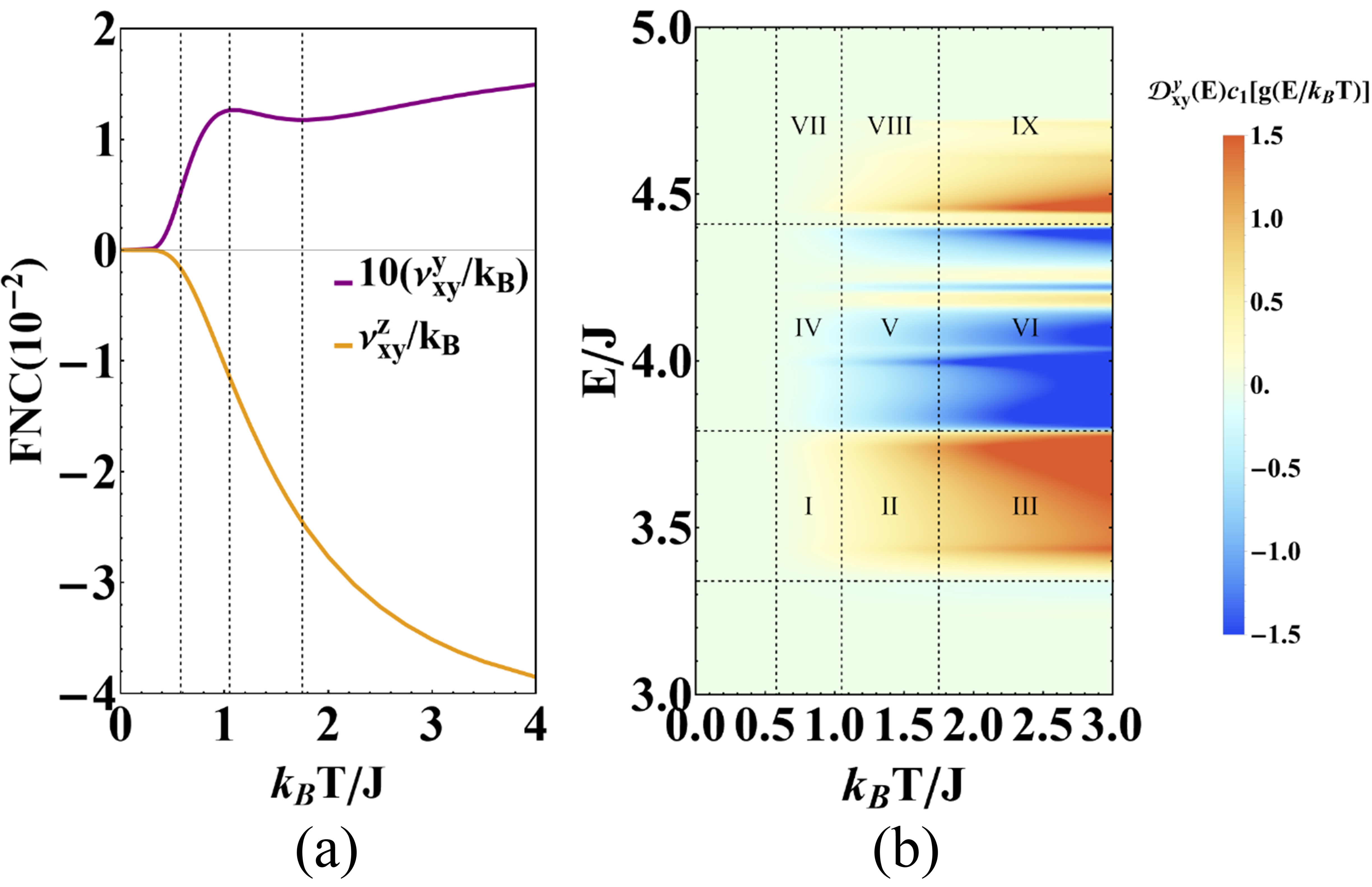}
    \caption{(a) Flavor Nernst coefficients $\nu^y_{xy}$ and $\nu^z_{xy}$ varies with $T$. We set $\eta/J=6$, $D/J=0.06$ and $B/J=0.6$. The $\nu^y_{xy}$ is one order smaller than $\nu^z_{xy}$. 
    (b) The $\mathcal{D}^y_{xy}(E) c_1[g(E/k_B T)]$ varies with $k_B T/J$ and $E/J$. The vertical dashed lines in both (a) and (b) are $k_B T/J=0.58$, $k_B T/J=1.05$ and $k_B T/J=1.75$. 
    The horizontal dashed lines in (b) divide the positive area and negative area of DOsB $\mathcal{D}^y_{xy}(E)$.}
    \label{fig:coe_with_T}
\end{figure}

In the quantum paramagnetic phase, with the linear flavor representation, one can obtain the flavor excitations by diagonalizing the BdG Hamiltonian Eq.~\eqref{eq:Ham}. The analytical expression for $A_{{\bm{k}}}$ and $B_{{\bm{k}}}$ can be found in the appendix~\ref{appendix:Bogoliubov–de Gennes Hamiltonian}. As an example, we depict the linear flavor wave dispersion in Fig.~\ref{fig:dis_pyro} with $\eta/J = 6$, $D/J=0.06$ and $B/J=0.6$. The bands are fully gapped by the anisotropy $\eta>0$ and the degeneracy from the time-reversal symmetry between opposite flavors is broken by the external magnetic field. Besides, compared to the 3D bulk pyrochlore lattice, the cubic symmetry is broken in the thin film case, resulting in 10 fully separated bands.

Before calculating the Nernst coefficients, one can notice that the flavor Nernst coefficient (FNC) tensor $\nu^s_{\alpha\beta}$ transforms as~\cite{suzuki2017cluster,mook2019thermal,li2020intrinsic}
\begin{align}
   \nu^s_{\alpha\beta} = \det(R) R_{s s'}R_{\alpha\alpha'}R_{\beta\beta'} \nu^{s'}_{\alpha'\beta'}
\end{align}
under $R$, the matrix representation of a symmetry element $\mathcal{R}$. Thus, the coefficient tensor is constrained by the lattice symmetry.

\begin{figure*}
    \centering
    \includegraphics[width=\textwidth]{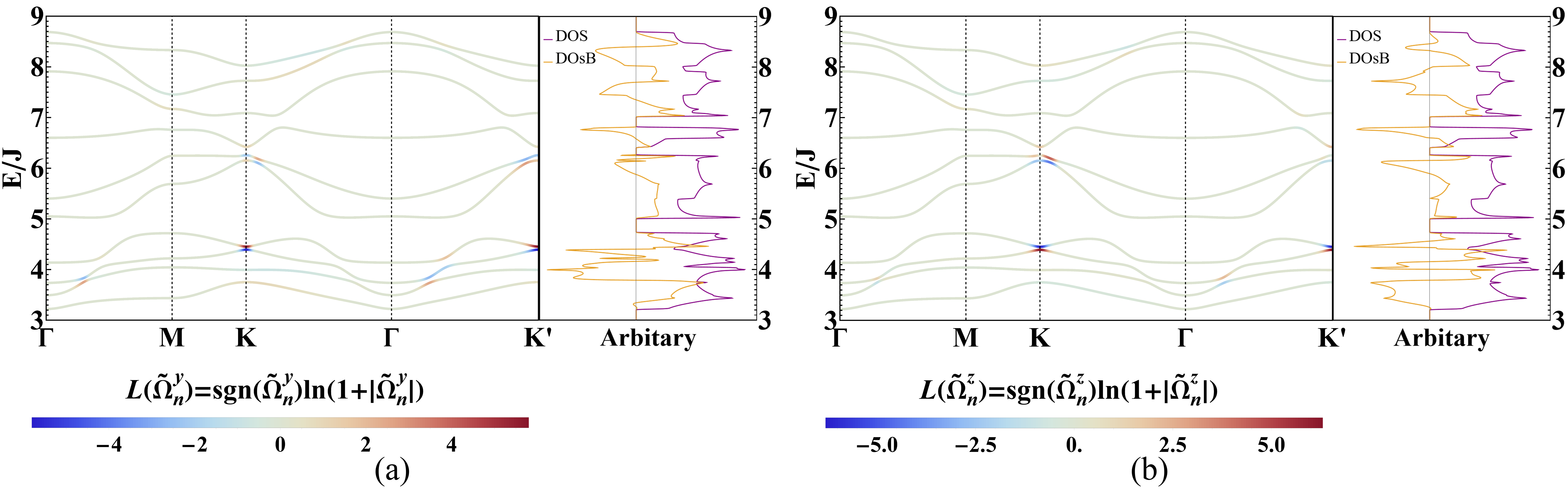}
    \caption{Dispersion colored by spin Berry curvature in Logarithm scale $L(\tilde{\Omega}^s_n)$ and the DOS $\mathcal{D}(E)$ and DOsB $\mathcal{D}^s_{\alpha \beta}(E)$. (a) is the situation for $s=y$ and (b) is the situation for $s=z$.}
    \label{fig:colored dispersion and DOsB}
\end{figure*}

Similar to the case in magnon Nernst effects of kagome lattices~\cite{li2020intrinsic}, two symmetries of the thin film pyrochlore, the mirror reflection about $yz$-plane plus time-reversal $\mathcal{M}_{yz} \mathcal{T}$, and the threefold rotation about $z$ axis $\mathcal{C}_{3z}$, require the coefficient tensors to be
\begin{align}
[\nu^x\!,\nu^y\!,\nu^z]=\begin{bmatrix}
    \begin{pmatrix}
    -\nu^y_{xy}     & 0\\
    0                  & \nu^y_{xy}\\
\end{pmatrix}\!,\begin{pmatrix}
    0                  & \nu^y_{xy}\\
    \nu^{y}_{xy}    & 0
\end{pmatrix}\!\begin{pmatrix}
    0                  & \nu^z_{xy}\\
    -\nu^z_{xy}      & 0
\end{pmatrix}
\end{bmatrix}.
\end{align}
Thus, there are only two independent coefficients $\nu^y_{xy}$ and $\nu^z_{xy}$ in this 2D pyrochlore thin film. It should also be noted that, although the AIAO Ising axis configuration classically gives rise to $\sum_i\langle S_i^y \rangle=0$, the transverse transport of spin-$y$ is symmetrically allowed. 

With the same parameter choice for Fig.~\ref{fig:dis_pyro}, we depict the two flavor Nernst coefficients $\nu^y_{xy}$ and $\nu^z_{xy}$ against temperature in Fig.~\ref{fig:coe_with_T}. We find that $\nu^y_{xy}$ is indeed non-zero but one order smaller than $\nu^z_{xy}$, reflecting the non-conservation of spin with small but finite $S^y_{n{\bm{k}}}=\langle\psi_{n{\bm{k}}}|S^y|\psi_{n{\bm{k}}}\rangle$ in the non-collinear Ising axis configuration.

Overall, the values of FNCs increase when the temperature rises, since the $c_1$-function in Eq.~\eqref{eq:nu} is larger at higher temperatures. Whereas the spin Berry curvature contribution from high-energy bands becomes more non-negligible when the temperature increases. This leads to the non-monotony of $\nu^y_{xy}$.

To understand the effect of the temperature in a more detailed manner, we color the band dispersion by $L(\tilde{\Omega}^s_{xy})=\text{sgn}(\tilde{\Omega}^s_{xy})\ln(1+|\tilde{\Omega}^s_{xy}|)$ in Fig.~\ref{fig:colored dispersion and DOsB}, where $L(x)=\text{sgn}(x)\ln{(1+|x|)}$ is Logarithmic function. It shows that non-zero spin Berry curvature mainly concentrates at anti-crossing points, and the spin Berry curvature of the two bands has opposite signs.

In Fig.~\ref{fig:colored dispersion and DOsB}, we also show the density of states (DOS) $\mathcal{D}(E)$ and the density of spin Berry curvature (DOsB) $\mathcal{D}^s_{\alpha\beta}(E)$, which are defined as
\begin{align}
    \mathcal{D}(E)&= \int_{BZ} \sum_n \frac{d{\bm{k}}}{(2\pi)^2} \delta(E - E_n({\bm{k}})),\\
    \mathcal{D}^s_{\alpha\beta}(E)&= \int_{BZ} \sum_n \frac{d{\bm{k}}}{(2\pi)^2} \delta(E - E_n({\bm{k}})) \tilde{\Omega}^s_{\alpha\beta,n{\bm{k}}}.
\end{align}
Since lower bands contribute more prominently to the FNE due to the boson statistics and the lowest four bands are energetically separated from those higher bands, we focus on these four bands, and one can clearly see the large DOsB with alternative signs below 4.8$J$, resulting in the non-monotonic behavior of $\nu^y_{xy}$.

From Fig.~\ref{fig:coe_with_T}(a), we find that $\nu^y_{xy}$ begins to increase significantly after reaching a certain temperature and reaches its maximum rate of increase at $k_B T/J=0.58$. As the temperature continues to rise, the coefficient reaches a peak at $k_B T/J=1.05$, followed by a decline, reaching a local minimum at $k_B T/J=1.75$, and then gradually increasing again. To explain the behavior of $\nu^y_{xy}$, we plot the $\mathcal{D}^y_{xy}(E) c_1[g(E/k_B T)]$ varies with $k_BT/J$ and $E/J$ in Fig.~\ref{fig:coe_with_T}(b). For $\mathcal{D}^y_{xy}(E)$, it's mainly positive (with red color) when $3.34 <E/J <3.79$ and it's mainly negative (with blue color) when $3.79< E/J <4.41$ which is showed in Fig.\ref{fig:colored dispersion and DOsB}. Therefore, we use these key data points to draw several regions with dashed lines in Fig.~\ref{fig:coe_with_T}(b). When $k_B T/J<0.58$, we do not observe the blue region at any energy level. When the temperature reaches the region $0.58< k_B T/J <1.05$, the negative region IV slows down the growth of $\nu^y_{xy}$. As the temperature gets to the region $1.05< k_B T/J <1.75$, the negative region V, especially the deep blue area around $E/J=4$, reverses the growth trend of the $\nu^y_{xy}$. As the temperature continues to increase, region IX and III become increasingly red and cause the $\nu^y_{xy}$ to resume its growth trend.

\begin{figure}
    \flushleft
    \includegraphics[width=0.45\textwidth]{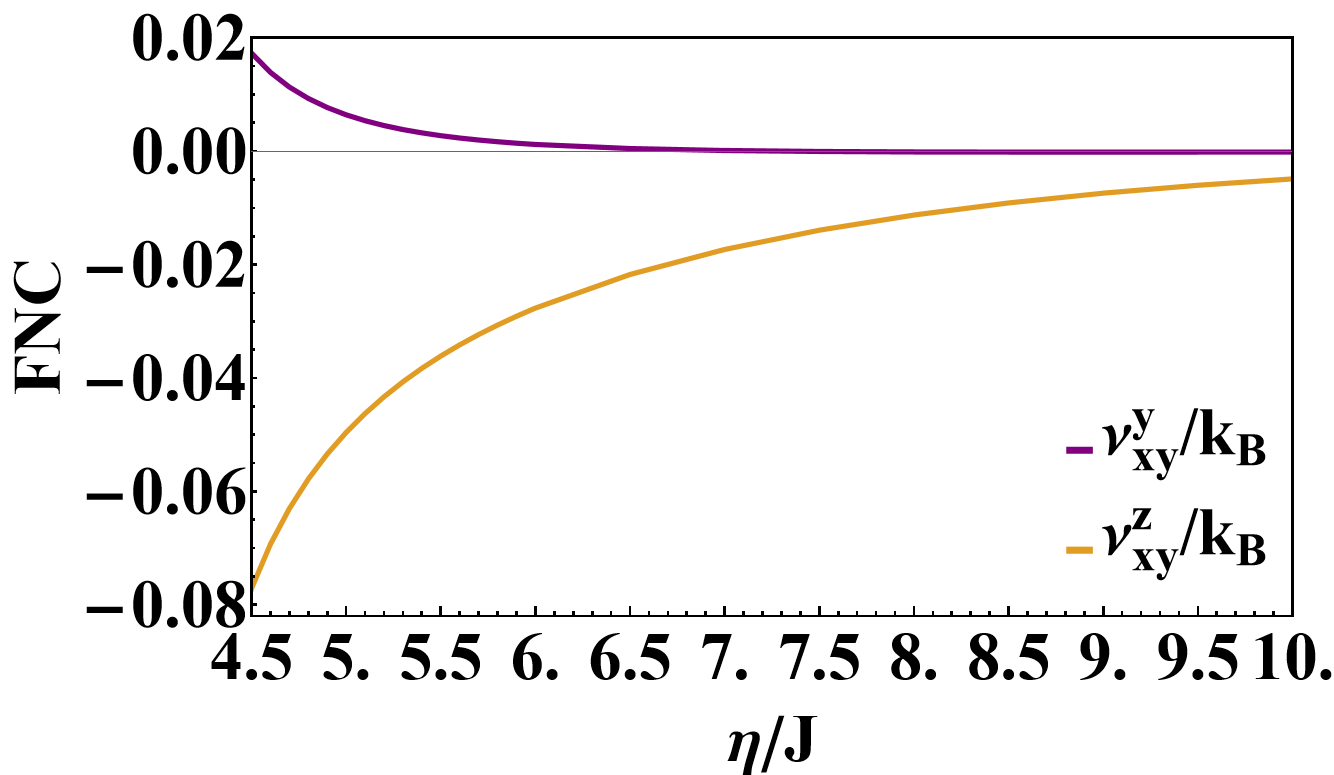}
    \caption{The flavor Nernst coefficients (in the unit of $k_B$) vary with $\eta$ when $k_B T/J=2$. Here $D/J=0.06$, $B/J=0.6$}
    \label{fig:D0.06_B0.6_eta}
\end{figure}
In addition to $k_BT$, another characterized energy scale is the anisotropy $\eta$, which determines the energy gap of the CEF excitations from the paramagnetic ground state. As shown in Fig.~\ref{fig:D0.06_B0.6_eta}, if $\eta$ is smaller, the CEF levels are easier to be thermally excited, giving rise to a larger thermal response. Taking the typical lattice constant $d\sim$ 10 \AA\  for pyrochlore materials~\cite{wen2021epitaxial} as an estimate for the thin film thickness, with a temperature gradient $\nabla_z T\sim$10 K/mm~\cite{lin2022Evidence} and the same parameter choice in Fig.~\ref{fig:D0.06_B0.6_eta}, one can obtain a experimentally accessible $z(y)$-polarized spin current up to 10$^{-11}$(10$^{-12}$) J/m$^2$.

\begin{figure}[!hbt]
    \flushleft
    \includegraphics[width=0.47\textwidth]{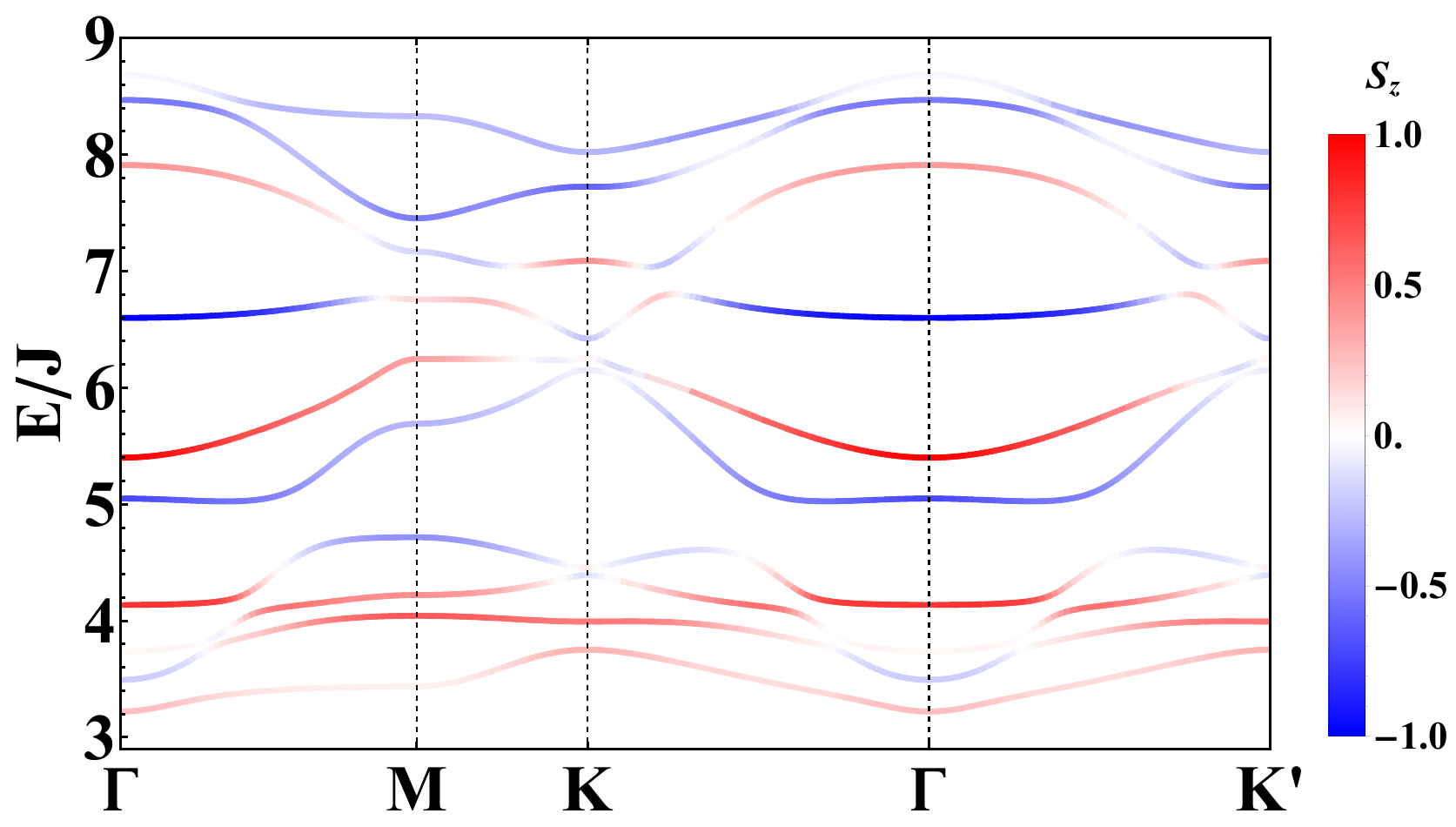}
    \caption{Dispersion colored by the momentum-resolved spin $z$-component $\bra{\psi_{n{\bm{k}}}}S^z\ket{\psi_{n{\bm{k}}}}$.}
    \label{fig:Sz-Dispersion}
\end{figure}

Obviously, besides the influence of energy scales, the spin Berry curvature of the bands itself can significantly affect the FNE. As we mentioned before, the spin Berry curvature is directly related to the Berry curvature in the collinear case, and thus the FNE should be reminiscent of the geometric structure of the bands even in the more general non-collinear cases. For example, if one reproduces the band dispersion colored by the momentum-resolved spin $z$-component $S^z_{n{\bm{k}}}\equiv\langle\psi_{n{\bm{k}}}|S^z|\psi_{n{\bm{k}}}\rangle$ as shown in Fig.~\ref{fig:Sz-Dispersion}, the lower three bands are mainly spin-up. Therefore, the spin Berry curvature is approximately connected with the band Berry curvature as $\tilde{\Omega}^z_{xy,n{\bm{k}}}\sim\left(\frac{1}{V}\sum_{\bm{k}}S^z_{n{\bm{k}}}\right)\Omega_{xy,n{\bm{k}}}$ such that the FNE indirectly reflects the geometric and topological properties of the bands.

\begin{figure}
    \flushleft
    \includegraphics[width=0.43\textwidth]{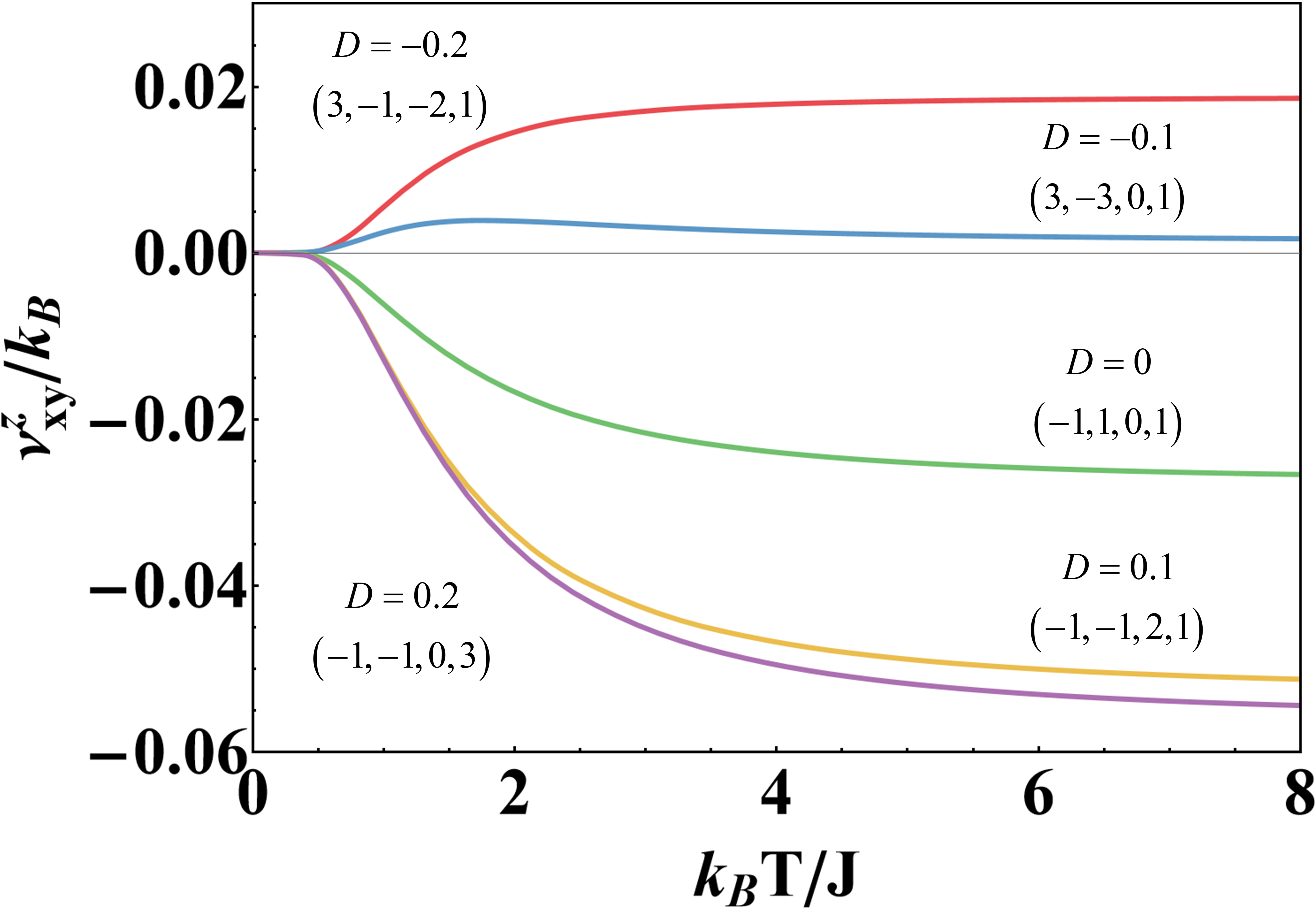}
    \caption{The flavor Nernst coefficients with varying parameters. We set $\eta/J=6$, $B/J=0.6$ with different $D/J$. We also label the corresponding Chern numbers of the lower four bands.}
    \label{fig:dmi_with_t}
\end{figure}

As the DM interaction can drastically change the band topology, we investigate the dependence of $\nu^{z}_{xy}$ on the temperature $T$ for different DM interactions $D$ in Fig.~\ref{fig:dmi_with_t}. 
For each curve, the four numbers from left to right indicate the Chern numbers of the lowest four bands from bottom to top, which match well with the behavior of the curve. For example, when ${D/J=-0.2}$ or $-0.1$, the lowest band has a positive Chern number $+3$, leading to a positive flavor Nernst coefficient. Meanwhile, comparing to the large positive response when $D/J=-0.2$, the large negative Chern number $-3$ of the second band in the case of $D/J=-0.1$ suppresses the increases of $\nu^z_{xy}$ when the temperature rises. In contrast, for $D/J=0$, $0.1$ and $0.2$, a negative flavor Nernst coefficient occurs due to the negative Chern number $-1$ of the lowest band. It should be noted that, owing to the non-collinearity of the Ising axes, the band Chern numbers as well as the flavor Nernst coefficients are non-zero even when $D=0$. Since a small deviation of the DMI can largely change the band topology as well as the thermal response, it is then possible to control the thermal spin current easily with strain engineering, especially in layered systems or heterostructures~\cite{kim2020StrainEngineeringMagnetic,zhang2021strain,xu2022strain}. In those cases, the strain applied along the layered direction can tune both the bulk or interfacial DM vectors and the direction of the Ising axes. Thus, our theory provides a potential route for manipulating thermal spin currents and may find its application in the field of piezo-spintronics~\cite{nunez2014theory,ulloa2017piezospintronic,liu2019antiferromagnetic}.

\begin{figure}[!ht]
    \flushleft
    \includegraphics[width=0.4\textwidth]{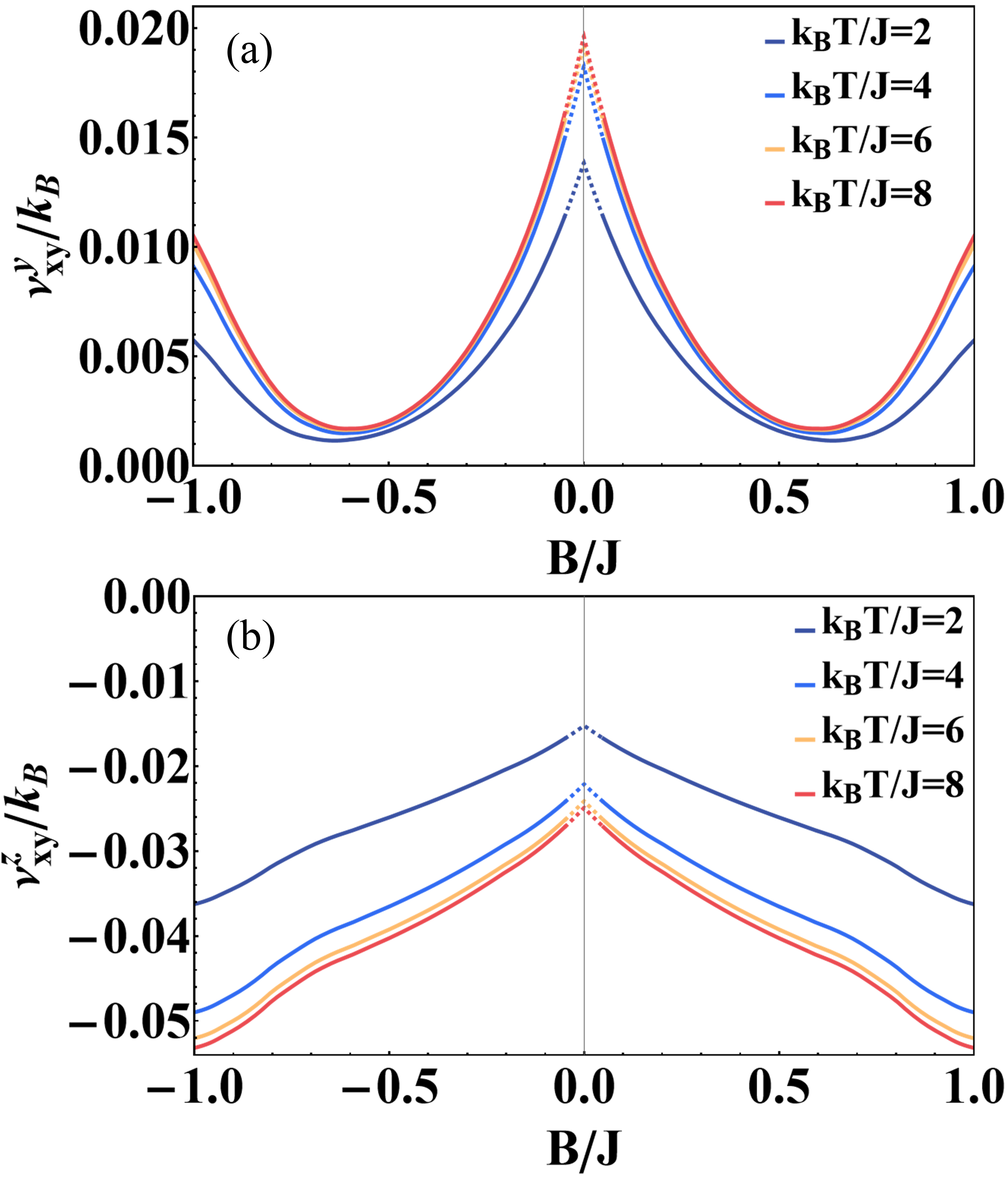}
    \caption{The flavor Nernst coefficients vary with magnetic field $B$ in different temperature $k_B T/J$. (a) is for $\nu^y_{xy}$ and (b) is for $\nu^z_{xy}$. We set $\eta/J=6$, $D/J=0.06$. The dashed lines are extrapolations from calculations.}
    \label{fig:Dz6_D0.06_B_T}
\end{figure}

Because the magnetic field can change the band dispersion as well as spin Berry curvature distribution, the FNE can also be tuned by the magnetic field.  But different from thermal Hall effects, reversing the magnetic field does not flip the thermal spin current response, due to the fact that the spin Berry curvature is symmetric under time reversal. In Fig~\ref{fig:Dz6_D0.06_B_T}, we show the dependence of $\nu^y_{xy}$ and $\nu^z_{xy}$ on the external magnetic field $B$ (in the unit of $J$) at different temperatures. We find that the response coefficients are indeed even functions of $B$, and thus the coefficients does not necessarily go to zero when $B\rightarrow 0$.

\section{Discussions}
\label{sec:Discussions}

While our numerical results are mainly for the 2D pyrochlore thin film, we believe that the underlying physics and predicted phenomena have broader relevance to related frustrated magnets in various lattices, such as honeycomb~\cite{ganesh2011quantum,joshi2018mathbb,liu2020featureless} or kagome lattices~\cite{ma2024upper}, with non-collinear Ising axes and/or non-zero DMI. Besides, our theory can readily extend to 3D bulk systems, where gapless topology and nodal line physics may naturally occur~\cite{li2018competing,li2016weyl,hwang2020topological}. Another intriguing possibility is the exploration of naturally occurring layered kagome systems, which share many frustration-induced properties with pyrochlore lattices. Materials such as herbertsmithite \ce{ZnCu3(OH)6Cl2} and kapellasite \ce{Cu3Zn(OH)6Cl2} have garnered significant attention as potential quantum spin liquid candidates~\cite{norman2016ColloquiumHerbertsmithite}. While not identical to our 2D pyrochlore model, these systems may exhibit similar flavor Nernst phenomena and could serve as valuable experimental testbeds for our predictions.

To capture the essential physics of the flavor Nernst effects in a straightforward way, our work is based on a simplified spin-1 Hamiltonian by only considering a singlet ground state and doubly degenerated first excited states. Although in real materials the CEF excitations with non-zero spin degrees of freedom can be much more complicated, the detailed information of the CEF levels can be investigated with the help of first-principle calculations~\cite{brooks1997density} or neutron scattering~\cite{frick1986crystal,zhang2014neutron} and Raman spectroscopies~\cite{schaack2007raman}, and we expect the theoretical framework of our study can be applied to those candidate materials without too many technique obstacles. From this point of view, the ligand environment plays a significant role in the formation of the CEF excitations, and strain tuning can rearrange the CEF levels~\cite{jensen1991rare,ishikawa2017reversed,pinho2021impact} and indirectly affect the flavor Nernst effects. In addition to this external static distortion, the intrinsic dynamical distortion of the lattice, i.e. phonons, can also naturally couple with the CEF levels. It has been shown that spin-lattice coupling can induce non-trivial band topology in ordered magnets~\cite{go2019topological,park2019topological,ma2022antiferromagnetic,ma2024chiralphononsinducedspin} and phonons can develop non-zero angular momentum by hybridizing with CEF excitations~\cite{lujan2024spin,chaudhary2023gianteffectivemagneticmoments}. Similar results may also occur in the quantum magnetic states and introduce novel thermal spin current responses, which we leave for future study.

In summary, we have theoretically demonstrated a flavor Nernst effect in a thin film pyrochlore lattice with paramagnetic ground states, specifically focusing on the 2D pyrochlore thin film model. The FNE emerges from the interplay between spin-orbit coupling, crystal electric field excitations, and spin transport, offering a unique platform for studying topological spin transport phenomena. Our findings demonstrate the non-zero flavor Nernst coefficients, which are temperature, anisotropy, DM interaction and magnetic field dependent. The Berry curvature of the CEF excitations plays a crucial role in determining the magnitude and sign of the FNE, suggesting potential applications in manipulating thermal spin currents and exploring topological spin transport phenomena in quantum paramagnets. Future experimental investigations in materials such as pyrochlore and kagome magnets are encouraged to verify our theoretical predictions and explore the intriguing physics of flavor Nernst effects in quantum paramagnets.

\begin{acknowledgments}
We thank the useful discussions with Xi Luo. This work is supported by the National Science Foundation of China with Grant No.~92065203 and No.~12174067, by the Ministry of Science and Technology of China with Grant No.~2021YFA1400300, and by the Fundamental Research Funds for the Central Universities, Peking University,
by the Collaborative Research Fund of Hong Kong with Grant No.~C6009-20G and C7012-21G, by the Guangdong-Hong Kong Joint Laboratory of Quantum Matter, the NSFC/RGC JRS grant with Grant No.~N$\_$HKU774/21, by the General Research Fund of Hong Kong with Grants No.~17310622 and No.~17303023.
\end{acknowledgments}

\appendix

\section{Bogoliubov–de Gennes Hamiltonian}
\label{appendix:Bogoliubov–de Gennes Hamiltonian}

In this section, we will give the explicit form of the Bogoliubov-de Gennes (BdG) Hamiltonian for the pyrochlore thin film Eq.~\eqref{eq:Full_Ham} in the main text.

\begin{figure}
    \centering
    \includegraphics[width=0.45\textwidth]{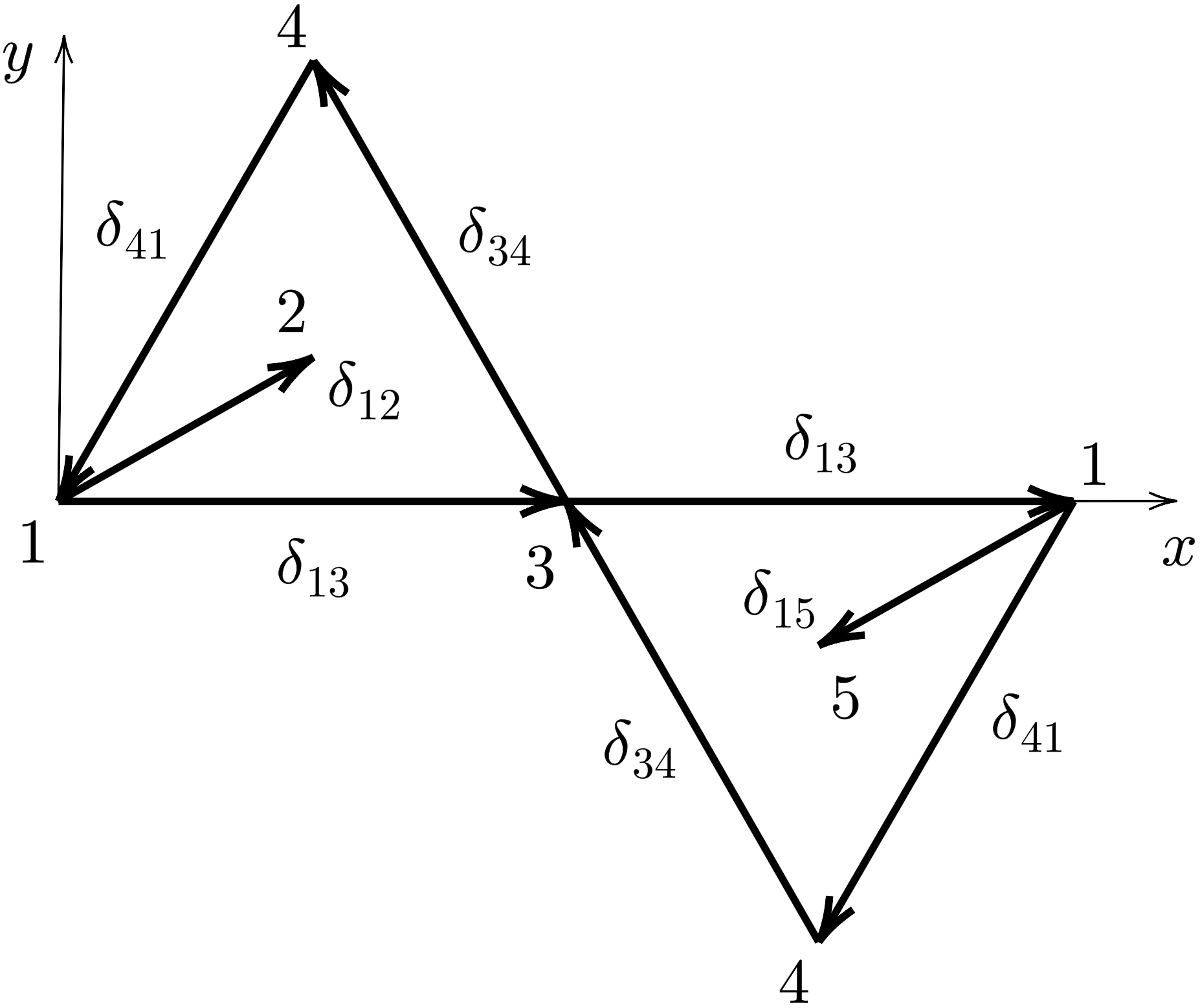}
    \caption{Position vectors diagram.}
    \label{fig:delta}
\end{figure}

The position vectors defined as $\bm{\delta}_{ij}=\bm{r}_j-\bm{r}_i$ are showed in Fig~\ref{fig:delta}, 
which are
\begin{gather}
    \bm{\delta}_{13}=(\frac{\sqrt{3}}{2},0,0),\  \bm{\delta}_{34}=(-\frac{\sqrt{3}}{4},\frac{3}{4},0),\\
    \bm{\delta}_{41}=(-\frac{\sqrt{3}}{4},-\frac{3}{4},0),\ \bm{\delta}_{12}=(\frac{\sqrt{3}}{4}
    \frac{1}{4},\frac{\sqrt{2}}{2}),\\
    \bm{\delta}_{32}=(-\frac{\sqrt{3}}{4},\frac{1}{4},\frac{\sqrt{2}}{2}),\  \bm{\delta}_{42}=(0,-\frac{1}{2},\frac{\sqrt{2}}{2}), \\
    \bm{\delta}_{i5}=-\bm{\delta}_{i2},\\
    \bm{\delta}_{ij}=-\bm{\delta}_{ji},
\end{gather}
where $\bm{r}_i$ is the position vector of i site.

By doing the Fourier transform, we can obtain the BdG Hamiltonian in momentum space as
\begin{align}
    H=\frac{1}{2}\sum_{{\bm{k}}}\bm{\Psi}_{{\bm{k}}}^\dagger H_{{\bm{k}}}\bm{\Psi}_{{\bm{k}}}
    =\frac{1}{2}\sum_{{\bm{k}}}\bm{\Psi}_{{\bm{k}}}^\dagger 
    \begin{pmatrix}
        A_{{\bm{k}}}    & B_{{\bm{k}}}\\
        B^*_{-{\bm{k}}} & A^*_{-{\bm{k}}}
    \end{pmatrix}
    \bm{\Psi}_{{\bm{k}}}
\end{align}
with
\begin{align}
    A_{ij}({\bm{k}}) =
    \begin{cases}
    -\bm{B} \cdot \bm{z}_{i} 
    \begin{pmatrix}
    1 & 0\\
    0 & -1
    \end{pmatrix} 
    +\eta I_2  &,i=j,\\
    0 I_2  &,\text{when } (i,j) \in \mathcal{S},\\
    m_{ij} C_{ij}({\bm{k}}) &,\text{others},
    \end{cases}
\end{align}
and
\begin{align}
    B_{ij}({\bm{k}}) =
    \begin{cases}
    0 I_{2} & ,i=j\quad \text{and } (i,j) \in \mathcal{S},\\
    n_{ij} C_{ij}({\bm{k}}) & ,\text{others},
    \end{cases}
\end{align}
where
\begin{gather}
    m_{ij} =\frac{1}{3}
    \begin{pmatrix}
    -\sqrt{2} D-J & \left( -\sqrt{2} D +2J\right) \gamma _{ij}\\
    \left( -\sqrt{2} D +2J\right) \gamma _{ij}^{*} & -\sqrt{2} D-J
    \end{pmatrix},\\
    n_{ij} =\frac{1}{3}
    \begin{pmatrix}
    \left( -\sqrt{2} D +2J\right) \gamma _{ij} & -\sqrt{2} D-J\\
    -\sqrt{2} D-J & \left( -\sqrt{2} D +2J\right) \gamma _{ij}^{*}
    \end{pmatrix},\\
    C_{ij}({\bm{k}}) =
    \begin{cases}
    2\cos({\bm{k}} \cdot \delta _{ij}) & ,\text{when }( i,j) \in \mathcal{P},\\
    e^{i{\bm{k}} \cdot \bm{\delta }_{ij}} & ,\text{others},
    \end{cases}\\
    \mathcal{S}=\{(2,5),(5,2)\},\\
    \mathcal{P}=\{( 1,3) ,( 1,4) ,( 3,4),(3,1),(4,1),(4,3)\}.
\end{gather}

To diagonalize the BdG Hamiltonian, a matrix $\mathcal{Q}$ transforms $\bm{\Psi}_{{\bm{k}}}$ to $\bm{\psi}_{{\bm{k}}}$ such that $\bm{\Psi}_{{\bm{k}}}=\mathcal{Q}_{{\bm{k}}}\bm{\psi}_{{\bm{k}}}$ and $\mathcal{Q}_{{\bm{k}}}^\dagger H_{{\bm{k}}} \mathcal{Q}^{}_{{\bm{k}}}=\mathcal{E}_{{\bm{k}}}$, where $\mathcal{E}_{{\bm{k}}}$ is a diagonal matrix whose elements are the eigen-energies, i.e.,
\begin{align}
    \mathcal{E}_{{\bm{k}}}=
    \begin{pmatrix}
        E _{1{\bm{k}}} & \cdots & 0 &   &   & \\
        \vdots & \ddots & \vdots & & &\\
        0 & \cdots & E _{2m{\bm{k}}} &   &  &\\
        & & & E_{1,-{\bm{k}}}&\cdots  & 0 \\
        & & & \vdots & \ddots  & \vdots\\
        & & & 0 &\cdots & E_{2m,-{\bm{k}}}
    \end{pmatrix}.
\end{align}
To preserve the bosonic commutator as $[\bm{\Psi}^{}_{{\bm{k}}},\bm{\Psi}^\dagger_{{\bm{k}}}]=[\bm{\psi}^{}_{{\bm{k}}},\bm{\psi}^\dagger_{{\bm{k}}}]=\Sigma_z$, one has
\begin{align}
    \Sigma_z=[\bm{\Psi}^{}_{{\bm{k}}},\bm{\Psi}^\dagger_{{\bm{k}}}]=\mathcal{Q}^{}_{{\bm{k}}}[\bm{\psi}^{}_{{\bm{k}}},\bm{\psi}^\dagger_{{\bm{k}}}] \mathcal{Q}^\dagger_{{\bm{k}}} = \mathcal{Q}_{{\bm{k}}} \Sigma_z \mathcal{Q}^\dagger_{{\bm{k}}},
    \label{eq:Normalize_Q}
\end{align}
which means that $\mathcal{Q}$ is a paraunitary matrix, and thus $\mathcal{Q}^\dagger_{{\bm{k}}} = \Sigma_z \mathcal{Q}_{{\bm{k}}}^{-1}\Sigma_z$. Then,
\begin{equation}
    \mathcal{Q}^\dagger_{{\bm{k}}} H_{{\bm{k}}} \mathcal{Q}^{}_{{\bm{k}}} = \mathcal{E}_{{\bm{k}}} \quad
    \Rightarrow \quad \mathcal{Q}^{-1}_{{\bm{k}}} \Sigma_z H_{{\bm{k}}}\mathcal{Q}_{{\bm{k}}}= \Sigma_z \mathcal{E}_{{\bm{k}}}.
\end{equation}
Therefore, one can obtain the eigen-energy of the BdG Hamiltonian by computing the positive eigenvalues of $\Sigma_z H_{{\bm{k}}}$.

\section{Spin Berry curvature with collinear Ising axes}
\label{appendix:Spin Berry curvature}
In the collinear case with the Ising axes along $z$-direction, spin $z$ is conserved. If no degeneracy, $S^z$ and $H_{\bm{k}}$ share the same eigenstates $\psi_{\bm{k}}$ such that $S^z |\psi_{n{\bm{k}}}\rangle=s_n^z|\psi_{n{\bm{k}}}\rangle$ and $ H_{\bm{k}}|\psi_{n{\bm{k}}}\rangle=E_{n{\bm{k}}}|\psi_{n{\bm{k}}}\rangle$. Noticing that $\sum_n|\psi_{n{\bm{k}}}\rangle\Sigma_z\langle\psi_{n{\bm{k}}}|=\Sigma_z$, one has
\begin{align}
    S^z\Sigma_z H_{\bm{k}}&=S^z\left(\sum_n|\psi_{n{\bm{k}}}\rangle\Sigma_z\langle\psi_{n{\bm{k}}}|\right) H_{\bm{k}}\nonumber\\
    &=\sum_n \left(s_n^z|\psi_{n{\bm{k}}}\rangle\Sigma_z\langle\psi_{n{\bm{k}}}| E_{n{\bm{k}}}\right)\nonumber\\
    &=\sum_n \left(E_{n{\bm{k}}}|\psi_{n{\bm{k}}}\rangle\Sigma_z\langle\psi_{n{\bm{k}}}|s_n^z\right)=H_{\bm{k}}\Sigma_z S^z,
\end{align}
so that $j_{\alpha {\bm{k}}}^{z}=\frac{1}{2} S^{z} \Sigma_z v_{\alpha {\bm{k}}}$, and thus
\begin{align}
        &\tilde{\Omega}_{\alpha \beta ,n{\bm{k}}}^{z} =\sum _{n'\neq n} (\Sigma _{z} )_{nn}\frac{2\text{Im} [(j_{\alpha {\bm{k}}}^{z} )_{nn'} (\Sigma _{z} )_{n'n'} (v_{\beta {\bm{k}}} )_{n'n} ]}{[ (\Sigma _{z} )_{nn} E_{n{\bm{k}}} -(\Sigma _{z} )_{n'n'} E_{n'{\bm{k}}}]^{2}}\nonumber\\
         & =\sum _{n'\neq n} (\Sigma _{z} )_{nn}\frac{2\text{Im} [(\frac{1}{2} S^{z} \Sigma_z v_{\alpha {\bm{k}}} )_{nn'} (\Sigma _{z} )_{n'n'} (v_{\beta {\bm{k}}} )_{n'n} ]}{[ (\Sigma _{z} )_{nn} E_{n{\bm{k}}} -(\Sigma _{z} )_{n'n'} E_{n'{\bm{k}}}]^{2}}\nonumber\\
         & =\frac{1}{2}\sum _{n'\neq n} (\Sigma _{z} )_{nn}\frac{2\text{Im} [(S^{z} \Sigma_z v_{\alpha {\bm{k}}} )_{nn'} (\Sigma _{z} )_{n'n'} (v_{\beta {\bm{k}}} )_{n'n} ]}{[ (\Sigma _{z} )_{nn} E_{n{\bm{k}}} -(\Sigma _{z} )_{n'n'} E_{n'{\bm{k}}}]^{2}}\nonumber\\
         & =\frac{1}{2}\sum _{n'\neq n} (\Sigma _{z} )_{nn}\frac{2\text{Im} [(S^{z} \Sigma_z)_{nm}( v_{\alpha {\bm{k}}} )_{mn'} (\Sigma _{z} )_{n'n'} (v_{\beta {\bm{k}}} )_{n'n} ]}{[ (\Sigma _{z} )_{nn} E_{n{\bm{k}}} -(\Sigma _{z} )_{n'n'} E_{n'{\bm{k}}}]^{2}}\nonumber\\
         & =\frac{1}{2} (S^{z} \Sigma_z)_{nn}\sum _{n'\neq n} (\Sigma _{z} )_{nn}\frac{2\text{Im} [( v_{\alpha {\bm{k}}} )_{nn'} (\Sigma _{z} )_{n'n'} (v_{\beta {\bm{k}}} )_{n'n} ]}{[ (\Sigma _{z} )_{nn} E_{n{\bm{k}}} -(\Sigma _{z} )_{n'n'} E_{n'{\bm{k}}}]^{2}}\nonumber\\
         & =\frac{1}{2}s_n^{z} \Omega_{\alpha \beta ,n{\bm{k}}},
    \end{align}
where we used the condition that $S^z\Sigma_z$ is diagonal in the eigen-bases $|\psi_{n{\bm{k}}}\rangle$ and $(\Sigma_z)_{nn}=+1$ for particle bands. 

\begin{figure}
    \centering
    \includegraphics[width=0.48\textwidth]{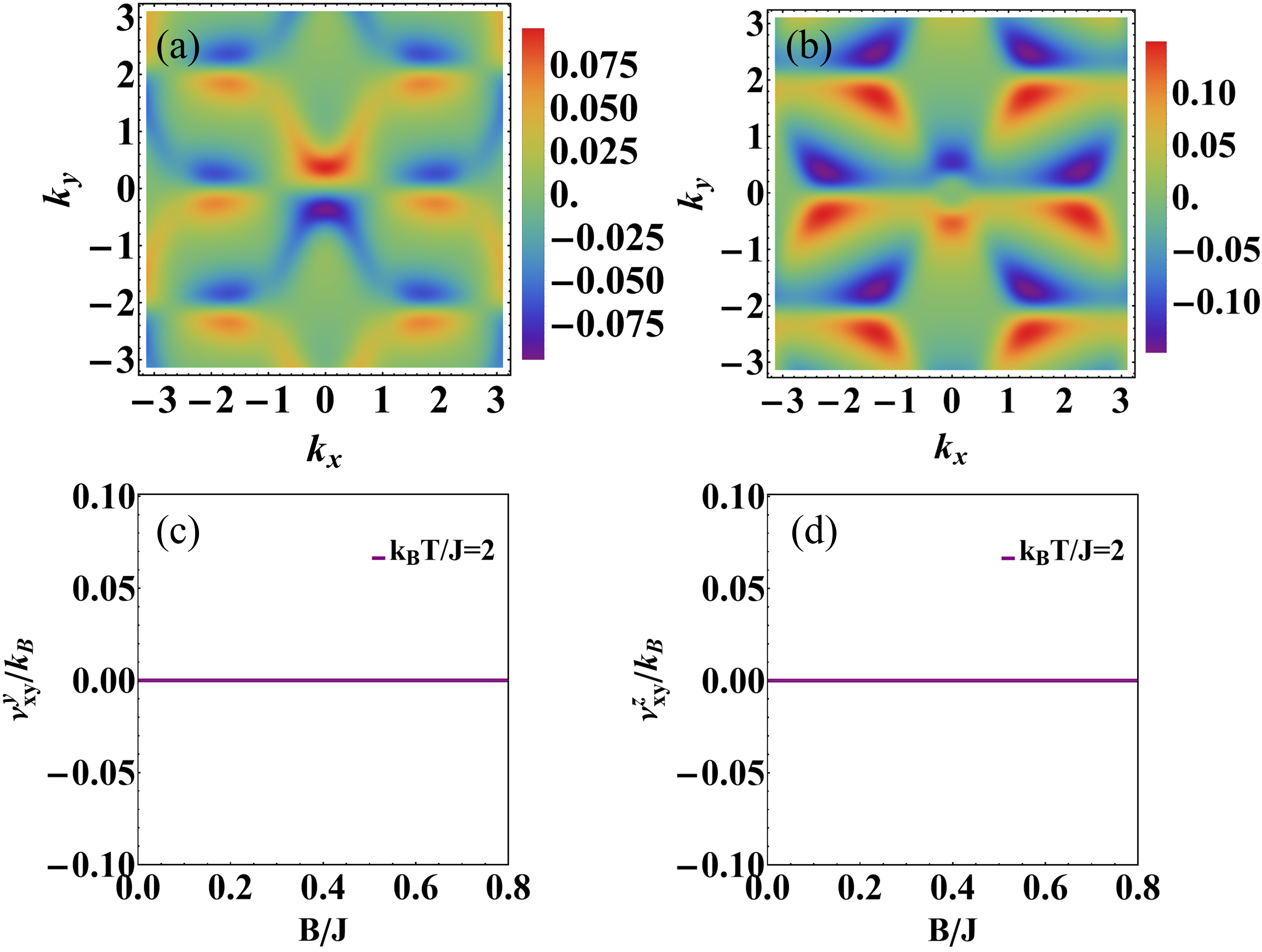}
    \caption{We set $\eta/J=6, D/J=0.06, B/J=0.6$. (a) The generalized Berry curvature of the torque when $s=y$. Here is the diagram for first band. (b) The generalized Berry curvature of the torque when $s=z$ for the first band. (c) and (d) are the coefficients of $\nu^y_{xy}$ and $\nu^z_{xy}$. They are indeed zero.}
    \label{fig:Torque}
\end{figure}
\section{Zero Torque response}
In this section, we show that the thermal response of the torque term in Eq.~\eqref{eq:Heisenberg equation} is indeed zero, and thus the thermal spin current remains to be well-defined even spin U(1) rotational symmetry is broken.

In Eq.~\eqref{eq:Heisenberg equation}, the torque term is 
\begin{equation}
    \tilde{T}^s=-\frac{i}{2}\sum_{\bm{k}}\bm{\Psi}_{\bm{k}}^\dagger\left[S^s\Sigma_z H_{\bm{k}}-H_{\bm{k}}\Sigma_z S^s\right]\bm{\Psi}_{\bm{k}}.
\end{equation}
Similar to Eq.~\eqref{eq:nu}, we can calculate the torque response induced from heat current by replacing $\hat{\bm{j}}^s$ with $\hat{T}^s$. By linear response theory, the torque response coefficient $(\nu_T)^s_y=T^s/\nabla_y T$ can be expressed as
\begin{align}
    (\nu_T)^s_{\alpha}=\frac{2k_B}{V}\sum_{n=1}^{2m}\sum_{{\bm{k}}}(\tilde{\Omega}_T)^s_{\alpha\beta,n{\bm{k}}}c_1[g(E_{n{\bm{k}}})],
    \label{eq:nu_T}
\end{align}
with
$(\tilde{\Omega}_T)^s_{\alpha,n{\bm{k}}}$ defined as
\begin{equation}
    (\tilde{\Omega}_T)^s_{\alpha,n{\bm{k}}}=\sum_{n'\neq n}(\Sigma_z)_{nn}\frac{2\text{Im}[(T^s_{{\bm{k}}})_{nn'}(\Sigma_z)_{n'n'}(v_{\beta{\bm{k}}})_{n'n}]}{\left[(\Sigma_z)_{nn}E_{n{\bm{k}}}-(\Sigma_z)_{n'n'}E_{n'{\bm{k}}}\right]^2}.
    \label{GsBC_T}
\end{equation}

Figure \ref{fig:Torque}(a) and \ref{fig:Torque}(b) show the generalized Berry curvature of the torque term $(\tilde{\Omega}_T)^s_{\alpha,n{\bm{k}}}$ is indeed satisfied $(\tilde{\Omega}_T)^s_{\alpha,n{\bm{k}}}=-(\tilde{\Omega}_T)^s_{\alpha,n\bm{-k}}$. Figure \ref{fig:Torque}(c) and \ref{fig:Torque}(d) show the coefficient is zero.

\bibliography{FNE}
\end{document}